\documentclass[5p,twocolumn]{elsarticle}
\usepackage{graphicx,latexsym}
\usepackage{dcolumn}
\usepackage{amssymb,amsmath,bm}
\usepackage{subfigure}
\usepackage{braket}
\usepackage{siunitx}
\usepackage{array}
\usepackage{float}
\usepackage[nopar]{lipsum}
\usepackage{caption}
\usepackage{multirow}
\usepackage{bigstrut}

\biboptions{sort&compress}

\usepackage[super]{nth}

\newcommand{\angstrom}{\text{\normalfont\AA}}
\usepackage{hyperref}
\hypersetup{
    pdfnewwindow=true,       
    colorlinks=true,         
    linkcolor=blue,          
    citecolor=blue,          
    filecolor=magenta,       
    urlcolor=black           
}

\usepackage[normalem]{ulem}

\def\sec#1{Sec.\ \ref{#1}}

\def\fig#1{Fig.\ \ref{#1}}
\def\tab#1{Tab.\ \ref{#1}}

\journal{}

\begin{document}

\begin{frontmatter}


\title{Role of planar buckling on the electronic, thermal, and optical properties of Germagraphene nanosheets}

\author[a1,a2]{Nzar Rauf Abdullah}
\ead{nzar.r.abdullah@gmail.com}
\address[a1]{Division of Computational Nanoscience, Physics Department, College of Science,
             \\ University of Sulaimani, Sulaimani 46001, Kurdistan Region, Iraq.}
\address[a2]{Computer Engineering Department, College of Engineering,
	\\ Komar University of Science and Technology, Sulaimani 46001, Kurdistan Region, Iraq.}

\author[a3]{Yousif Hussein Azeez}
\address[a3]{Physics Department, College of Science, University of Halabja, Kurdistan Region, Iraq.}

\author[a4]{Botan Jawdat Abdullah}
\address[a4]{Physics Department, College of Science, Salahaddin University-Erbil, Erbil 44001, Kurdistan Region, Iraq.}

\author[a1]{Hunar Omar Rashid}

\author[a5]{Andrei Manolescu}
\address[a5]{Reykjavik University, School of Science and Engineering, Menntavegur 1, IS-101 Reykjavik, Iceland.}

\author[a6]{Vidar Gudmundsson}
\address[a6]{Science Institute, University of Iceland, Dunhaga 3, IS-107 Reykjavik, Iceland.}


\begin{abstract}

We report the electronic, the thermal, and the optical properties of a Germagraphene (GeC) monolayer taking into account buckling effects. The relatively wide direct band gap of a flat GeC nanosheet can be changed by tuning the planar buckling. A GeC monolayer has an sp$^2$ hybridization in which the contribution of an $s$-orbital is half of the contribution of a $p$-orbital leading to stronger $\sigma\text{-}\sigma$ bonds compared to the $\sigma\text{-}\pi$ bonds. Increasing the planar buckling, the contribution of an $s$-orbital is decreased while the contribution of a $p$-orbital is increased resulting in a sp$^3$-hybridization in which the $\sigma\text{-}\pi$ bond becomes stronger than the $\sigma\text{-}\sigma$ bond. As a result, the band gap of a buckled GeC is reduced and thus the thermal and the optical properties are significantly modified.
We find that the heat capacity of the buckled GeC is decreased at low values of planar buckling, which is caused by the anticrossing of the optical and the acoustic phonon modes affecting phonon scattering processes.
The resulting optical properties, such as the dielectric function, the refractive index, the electron energy loss spectra, the absorption, and the optical conductivity show that a buckled GeC nanosheet has increased optical activities in the visible light region compared to a flat GeC. The optical conductivity is red shifted from the near ultraviolet to the visible light region, when the planar buckling is increased. We can thus confirm that the buckling can be seen as another parameter to improve GeC monolayers for optoelectronic devices.

\end{abstract}


\end{frontmatter}

\section{Introduction}
Research on two-dimensional (2D) materials indicates a great potential for next-generation electronic and optical applications due to their rich physical characteristics and outstanding electronic properties \cite{Lemme2022, PhysRevLett.105.136805, ABDULLAH2021106981, C7RA02198D, C8RA08008A}.
However, some of the 2D materials such as graphene and silicene have a vanishing gap and are thus called
gapless materials. The vanishing gap causes problems for applications using graphene-based electronic devices. Thus, many investigations have tried to search for other 2D materials \cite{Akinwande2019, PhysRevB.90.085424}. In recent years, there a lot of attention has been given to new 2D materials such as BN \cite{C8NR05027A}, MoS$_2$ \cite{Kumar_2022}, BeO \cite{ABDULLAH2022107102, ABDULLAH2022106409}, and GeC \cite{C7TA10118J}, which have a wider band gap and can be considered as semiconductor materials.

Theoretical investigations have reported that GeC monolayers are semiconductors and structurally stable \cite{SUZUKI20102820, PhysRevB.92.075435}. This has led researchers to study them intensively.
In addition to computational analysis using density functional theory, experimental synthesis has been used to investigate the production of GeC monolayer as a possible 2D material. Various synthesis techniques, including plasma-enhanced chemical vapor deposition, activated reactive evaporation, and chemical vapor deposition can all be used to create germagraphene monolayers \cite{kumar1990thin, jamali2016effect, gupta2015synthesis}.

The band gap of a GeC monolayer is found to be $2.1$ eV (GGA) and $4.06$ eV using Heyd–Scuseria–Ernzerhof (HSE) hybrid functional at zero value for the buckling factor, i.e.\ a flat structure \cite{doi:10.1080/14786435.2016.1248517}. The valuable band gap of GeC indicating semiconducting properties can be further improved using several techniques in order to enhance its
possible role in thermoelectric and optoelectronic applications.
One may control the band gap of a fully hydrogenated GeC monolayer by biaxial strain or external
electric field and a semiconductor-metal phase transition takes place at certain elongation caused by
biaxial strain. The band gap has thus been enhanced to $3.49$ eV displaying photocatalytic characteristics for water  splitting \cite{VU2020113857}.  Likewise, the mechanical, electronic, and
magnetic properties of a GeC monolayer can be modified through hydrogen or halogen passivation \cite{SOHBATZADEH201888, DRISSI2016148}
Doping a GeC monolayer could be considering as another technique to modify the band gap. For instance,  F and C dopant atoms in a GeC monolayer disrupt
the planar structure and a surface-functionalized GeC monolayer with low-buckling results  \cite{VU2020106359}. With this type of doping the band gap is seen to vary from $2.8$~eV to $3.2$~eV in calculations using HSE.

In this work, we perform DFT calculations based on the Kohn-Sham formalism implemented in the Quantum espresso software package \cite{Giannozzi_2009, giannozzi2017advanced}. In the calculations, we tune the buckling parameter to study the electronic, the thermal, and the optical properties of a GeC monolayer. The results show that the buckling effects can be considered as an alternative way for controlling it's physical properties, such as the band gap, the thermal conductivity and the heat capacity.

The structure of the paper is as follows: \sec{Computational} includes details of the computational methods, and \sec{Results} demonstrates the calculated electrical, the thermal, and the optical properties for a GeC monolayer with different degree of buckling. The last section, \sec{conclusion},
is the conclusion.

\section{Methodology}\label{Computational}
A $2\times 2$ supercell of a GeC monolayer with equal number of Ge and C atoms is considered.
The GeC structure is fully relaxed with high values of cutoffs for the plane-waves kinetic energy and the charge densities fixed at $1088.5$~eV, and $1.088 \times 10^{4}$~eV, respectively \cite{ABDULLAH2022114705}. In the relaxation process, the
forces on the atoms are less than $10^{-5}$ eV/$\angstrom$, where a dense Monkhorst-Pack grid with $18 \times 18 \times 1$ is used.
The distance between GeC monolayers is assumed to be $20 \, \angstrom$ in the $z$-direction, which is long enough to cancel out interlayer interactions.
The generalized gradient approximation (GGA) is used with the Perdew-Burke-Ernzerhof (PBE)
functionals approximating the exchange and the correlation terms implemented in QE software \cite{ABDULLAH2022115554}.
In the calculations of the band structure and the density of states (DOS), Self-Consistent Field (SCF) and non-self-consistent field (NSCF) calculations are performed, respectively.
In these calculations, we use a Monkhorst-Pack grid of $18\times18\times1$ for the SCF  and $100 \times 100 \times 1$ for the NSCF \cite{ABDULLAH2022114590}.
The optical properties of a GeC monolayer are obtained using QE with the optical broadening of $0.1$~eV.

An ab initio molecular dynamics, AIMD, calculations are utilized to check the ther­modynamic stability. The calculations, done in the NVT ensemble, are performed for $10$~ps with a time step of $1.0$~fs using the heat bath approach described by Nosé-Hoover \cite{doi:10.1063/1.463940}.

The optical characteristics of the GeC monolayer can be calculated using a large number of empty bands which is taken into account to evaluate the dielectric properties, $\varepsilon(\omega) = \varepsilon_1(\omega) + i\varepsilon_2(\omega)$, where $\varepsilon_1$ and $\varepsilon_2$ are the real and the imaginary parts of the dielectric function. In the long wavelength limit $q \rightarrow 0$, $\varepsilon_2(\omega)$ is given in Refs.\ \cite{ZHANG2021149644, GOUDARZI2019130}
\begin{equation}
	\varepsilon_2 (\omega) = \frac{2 e^2 \pi}{\omega \varepsilon_0}
	\sum_{K, c, v} | \bra{\Psi_K^{v}} \vec{u} \cdot \vec{r} \ket{\Psi_K^c} |^2 \, \delta(E_K^c - E_K^v - \omega).
\end{equation}
Herein, $\omega$ indicates the frequency of the electromagnetic waves, $\varepsilon_0$ refers to the free space permittivity. The labels $v$ and $c$ indicate the valence and conduction bands, respectively, and $\vec{u}$ and $\vec{r}$ demonstrate the polarization and the position vectors of the electromagnetic field, respectively. The real part and the imaginary part of the complex dielectric functions are connected to each other by the Kramers-Kronig relation \cite{dressel2002electrodynamics, Zhang_2021}. Both $\varepsilon_1(\omega)$ and $\varepsilon_2(\omega)$ are obtained from the QE package. Once the dielectric functions are obtained, the real part of the refractive index is calculated as \cite{Mistrik2017}
\begin{equation}
	n(\omega) = \frac{1}{\sqrt{2}} \Bigg(  \Big[ \varepsilon_1^2(\omega) + \varepsilon_2^2(\omega)    \Big]^{\frac{1}{2}} + \varepsilon_1(\omega) \Bigg)^{\frac{1}{2}}.
	\label{eq_n}
\end{equation}

\noindent The optical conductivity is then computed from

\begin{equation}
	\sigma_{\rm optical} = \frac{-i \, \omega}{4 \pi} \Big[  \varepsilon(\omega) - 1  \Big].
	\label{eq_sigma}
\end{equation}

\section{Results}\label{Results}
In this section, we show the obtained results for the electronic, the thermal and the optical properties of a GeC monolayer with different values for the planar buckling parameter, $\Delta$. In addition, for the sake of comparison, we recalculate the physical properties of a flat GeC monolayer, $\Delta = 0.0$, and use them as reference points to compare to.

\subsection{Electronic states}
\begin{figure}[htb]
	\centering
	\includegraphics[width=0.48\textwidth]{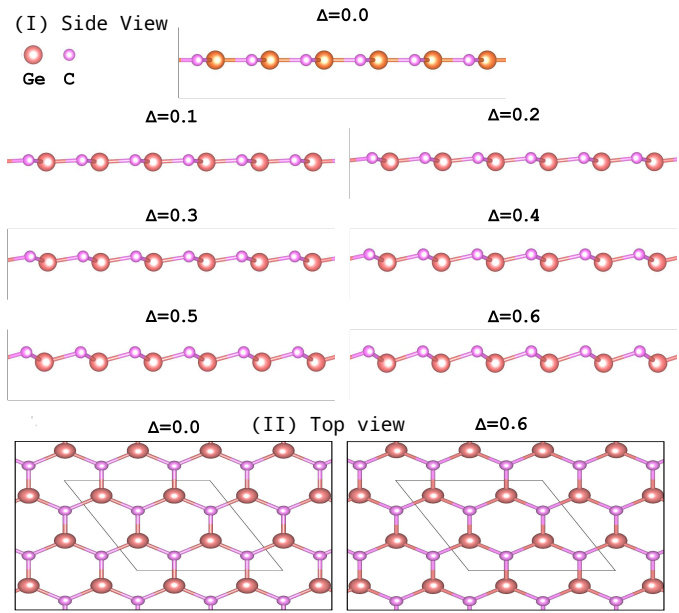}
	\caption{(I) Side view of crystal structure of a GeC monolayer for different values of planar buckling $\Delta$. (II) Top view of GeC monolayer for two values of planar buckling $\Delta = 0.0$ (left side) and $0.6$~$\angstrom$ (right side).}
	\label{fig01}
\end{figure}

In a flat, or planar, GeC monolayer ($\Delta = 0.0$), all the Ge and the C atoms are located in the same $xy$-plane as is presented in \fig{fig01} for both a side view (I), and a top view (II).
If a finite planar buckling is considered ($\Delta \neq 0.0$), the Ge atoms are located
in the same plane and all the C atoms are situated in another plane. The planar buckling indicates the vertical distance, $\Delta$, between the Ge and the C planes. There are several proposed techniques to control the planar buckling in monolayers experimentally. One of them is using an applied external electric field on the monolayer \cite{doi:10.1063/1.4979507}.

\begin{figure}[htb]
	\centering
	\includegraphics[width=0.43\textwidth]{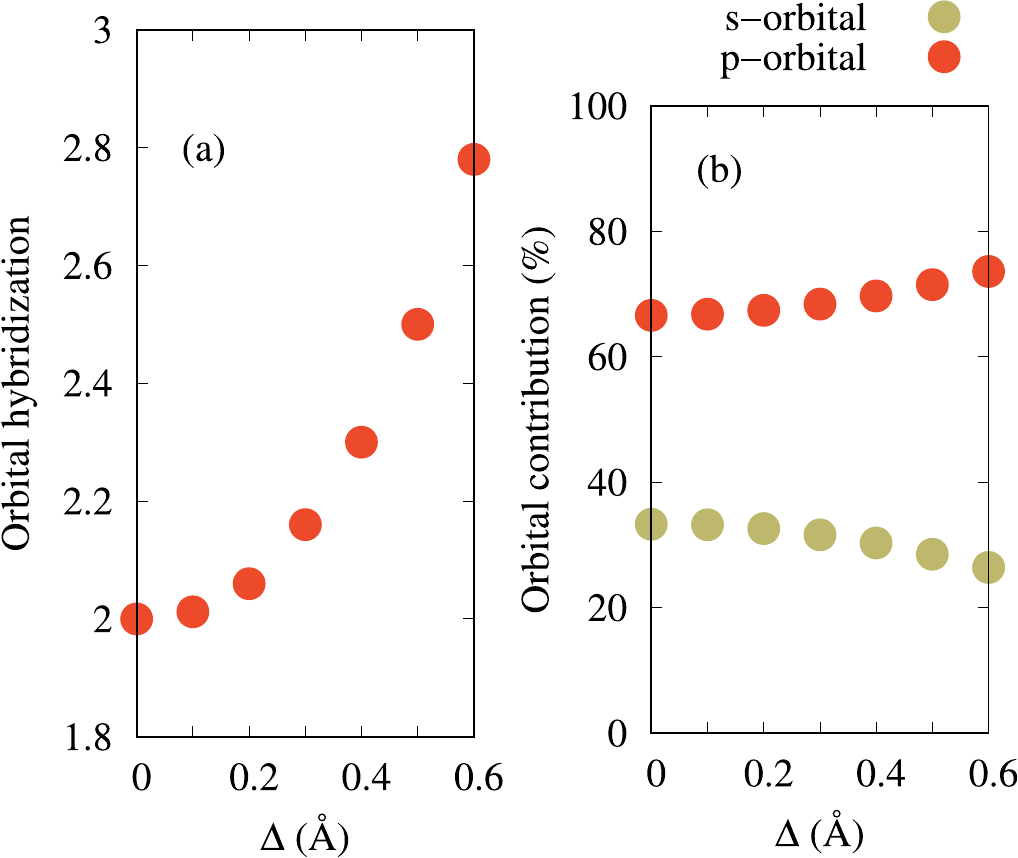}
	\caption{Orbital hybridization (a) and Orbital contribution (b) as a function of buckling $\Delta$ are shown.}
	\label{fig02}
\end{figure}

The degree of the hybridization of the $s$- and the $p$-orbitals can be found using a simple equation, $\cos(\theta) = s/(s-1) = (p-1)/p$, where $\theta$ is the angle between the equivalent orbitals, $s$ and $p$ \cite{kaufman1993inorganic}.
The considered values of planar buckling affect the orbital hybridization. Most flat monolayers have
an $sp^2$ hybridization, and the orbital hybridization is approaching an $sp^3$ configuration when the planar buckling is increased \cite{jalilian2016buckling, ABDULLAH2022106943}. A flat GeC monolayer ($\Delta = 0.0$) has an $sp^2$ hybridization as is presented in \fig{fig02}(a), where the degree of the orbital hybridization is $2$ indicating an $sp^2$-hybridization at $\Delta = 0.0$, and the degree of orbital hybridization is approaching $sp^3$ when the planar buckling is increased to $\Delta = 0.6$~$\angstrom$.
The maximum allowed value of planar buckling for a GeC monolayer is $0.6$~$\angstrom$ in which the orbital hybridization is $sp^{2.78}$. The orbital hybridization is correlated with the buckling parameter $\Delta$ in \tab{table_one} for all values considered.

\begin{table}[h]
	\centering
	\begin{center}
		\caption{\label{table_one} The formation energy (E$_{f}$), the bond length of Ge-C, the degree of hybridization for different values of planar buckling, $\Delta$.}
		\begin{tabular}{l|l|l|l}\hline
	   $\Delta (\angstrom)$	   &  Degree of hybridization   &   E$_f$ (eV)   &  Ge-C ($\angstrom$)  \\ \hline
			0.0	   &  sp$^2$        &    -       & 1.858          \\
			0.1	   &  sp$^{2.012}$  &  0.0665    & 1.861          \\
			0.2	   &  sp$^{2.06}$   &  0.2677    & 1.869          \\
			0.3	   &  sp$^{2.16}$   &  0.6111    & 1.882          \\
			0.4	   &  sp$^{2.3}$    &  1.1126    & 1.901          \\
			0.5	   &  sp$^{2.5}$    &  1.7992    & 1.924          \\
			0.6	   &  sp$^{2.78}$   &  2.7089    & 1.952          \\ \hline
	\end{tabular}	\end{center}
\end{table}

Increasing planar buckling leads to a change in the ratio between the $s$- and the $p$-orbital contribution.
For instance, the ratio of the $s$- and the $p$-orbitals are $33.333 \%$ and $66.666 \%$, respectively, for a flat GeC monolayer, that displays an $sp^2$-hybridization as is seen in \fig{fig02}(b).
Their ratio is changed to $ 26.4 \%$ and $ 73.6\%$ for a buckled GeC with $\Delta = 0.6$~$\angstrom$ indicating the role of the $s$-orbitals is decreases, while that of the the $p$-orbital increases, when the planar buckling is increased. This can be understood by the bonding character of the monolayer in which
strong $\sigma$ bonds are formed through the $sp^2$ orbital overlapping for a flat GeC, whereas the planar buckling reduces the $sp^2$ overlapping and the bond symmetry is broken simultaneously as it approaches an $sp^3$ overlapping. As a result, the $\pi$-bonds become stronger through the $p_z$-orbitals.

\begin{figure}[htb]
	\centering
	\includegraphics[width=0.45\textwidth]{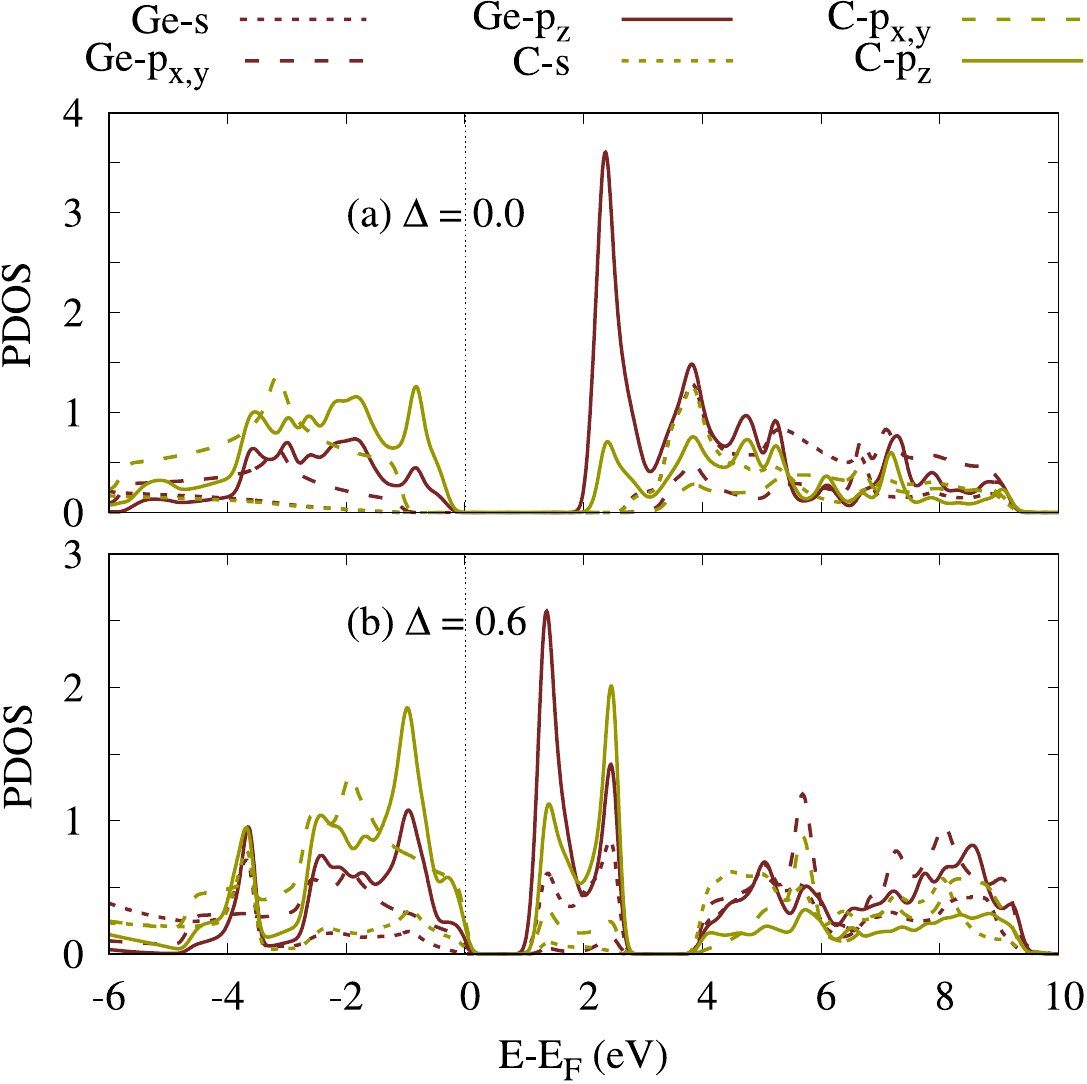}
	\caption{Partial density of states, PDOS, of the GeC monolayer without (a) and with buckling parameter, $\Delta = 0.6$~$\angstrom$ (b). The energies are with respect to the Fermi level, and the Fermi energy is set to zero}
	\label{fig03}
\end{figure}

To confirm the modification of the $\sigma$- and the $\pi$-bonds, we present the density of states of the $s$- and the $p$-orbitals (partial density of states, PDOS) for Ge and C atoms in \fig{fig03} for a flat GeC, $\Delta = 0.0$ (a), and a buckled GeC with $\Delta = 0.6$~$\angstrom$ (b). The orbital contribution of the C atoms is dominant in the valence band region (below the Fermi energy, E$_F$), while the Ge atoms have a higher orbital contribution in the conduction region.
It can be seen that the density of states of the $p_z$-orbitals is increased with planar buckling indicating that the $\pi_z$ bonds become stronger due to the planar buckling. On the other hand the density of states of the $s$-orbitals for the Ge atoms in the conduction region is decreased with increasing $\Delta$.

\begin{figure}[htb]
	\centering
	\includegraphics[width=0.48\textwidth]{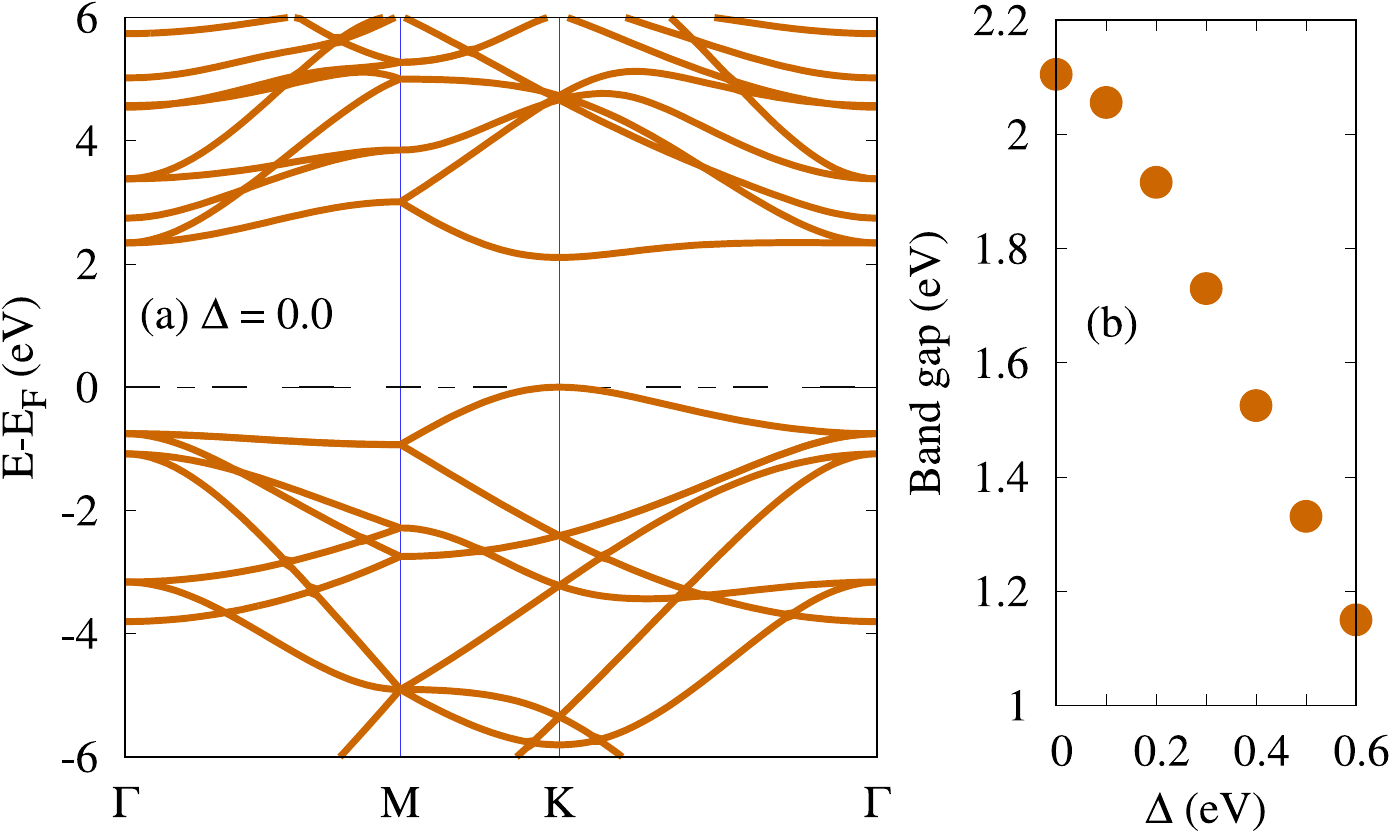}
	\caption{Band structure for optimized flat GeC monolayers with planer buckling, $\Delta = 0.0$ (a), and the band gap versus $\Delta$ (b) are plotted. The energies are with respect to the Fermi level, and the Fermi energy is set to zero.}
	\label{fig04}
\end{figure}

One can estimate the band gap of the GeC structure from the density of state in which the PDOS of both the $s$ and the $p$ orbitals of both atoms are shifted towards the Fermi energy resulting in a band gap reduction in the case of planar buckling. The band gap of a flat GeC is found to be $2.015$~eV, which is in a good agreement with the literature \cite{BEHZAD2019102306}. The band structure of a flat GeC monolayer is shown in \fig{fig04}(a) in addition to the band gap plotted as a function of the buckling parameter $\Delta$ in (b). The calculated electronic band structure
indicates that the flat GeC monolayer is a direct band gap semiconductor, with
both the valence band maximum and conduction band minimum located at the K-point.
The valence band maxima and the conduction band minima are composed of $\pi$ and $\pi^*$ bands which come from the C-2$p_z$ and the Ge-4$p_z$ orbitals extending above and below the GeC monolayer.

Increasing planar buckling reduces the band gap (see \fig{fig04}(b)) and the band gap is found to be $1.15$~eV at $\Delta = 0.6$~$\angstrom$. Furthermore, a direct to indirect band gap transition occurs when the planar buckling is considered, especially at the high value of planar buckling when $\Delta = 0.6$~$\angstrom$. The valence band maxima and the conduction band minima in the presence of the buckling
come from the combined C-2$p_{x,y}$, C-2$p_{z}$ and the Ge-4$p_z$ orbitals, respectively.

The formation energy, $E_f$, is an important parameter to determine whether a structure is stable or not.
The formation energy demonstrates the structural stability of the buckled GeC monolayers in comparison to a flat GeC monolayer. The $E_f$ values for a flat and the buckled GeC monolayers with different buckling
strength are listed in \tab{table_one}.
A GeC monolayer is stable for low values of $\Delta$, but
our calculations show that the formation energy increases with increasing $\Delta$.
As the planar buckling increases, a GeC monolayer becomes less energetically stable.
The same situation was seen for other monolayers, where the stability of a monolayer decreases with increasing planar buckling \cite{jalilian2016buckling}.
Additionally, the Ge-C bond length is elongated with increasing $\Delta$ (see \tab{table_one}). This leads to a symmetry breaking and thus affects the band structure and the band gap $E_g$, that is reduced with increasing Ge-C bond length.

\subsection{Thermal properties}
In this section, we present the phonon band structure and heat capacity of the GeC monolayers and show how the planar buckling affects their thermal stability and heat capacity.
\begin{figure}[htb]
	\centering
	\includegraphics[width=0.45\textwidth]{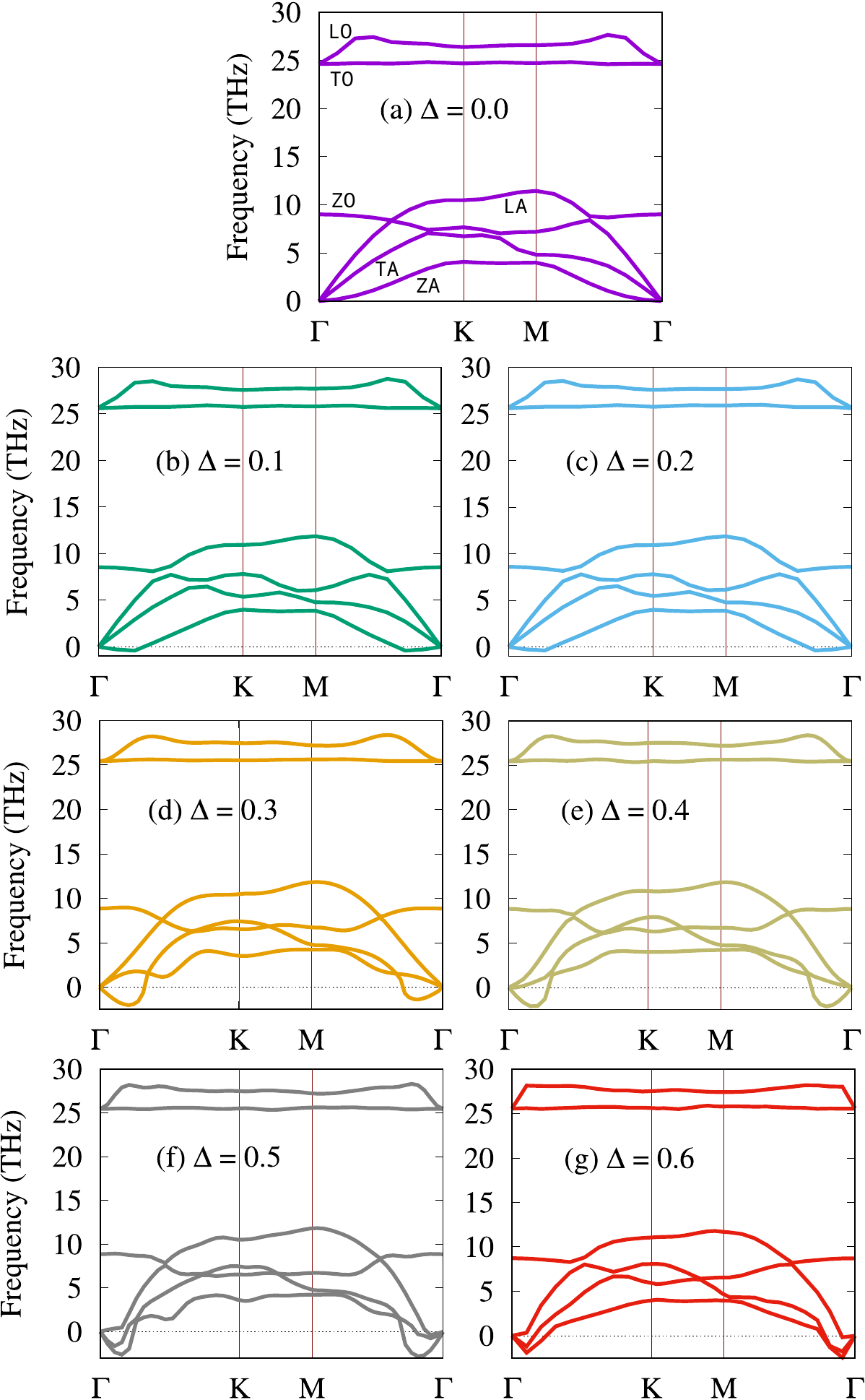}
	\caption{Phonon band structure of optimized GeC monolayers with planar buckling, $\Delta = 0.0$ (a), $0.1$ (b), $0.2$ (c), $0.3$ (d), $0.4$ (e), $0.5$ (f), and $0.6$~$\angstrom$ (g).}
	\label{fig05}
\end{figure}

We first take a look at the phonon band structure along the high symmetry directions shown in \fig{fig05}. The phonon band structure of a flat GeC monolayer, $\Delta = 0.0$, (\fig{fig05} a) has no imaginary frequencies indicating the dynamical stability of the monolayer. The phonon band structure is plotted for a unit cell of a GeC containing two atoms.
We thus see six phonon branches grouped into three acoustic and three optical ones. The acoustic phonon modes are in-plane longitudinal acoustic (LA), in-plane transverse acoustic (TA), and out-of-plane acoustic (ZA).
In the same way, the optical phonon modes are in-plane longitudinal optical (LO), in-plane transverse optical (TO), and out-of-plane optical (ZO). One can clearly see that the width of acoustic bands is much smaller than the phonon band gap (13.16 THz), which is important for the phonon transport.
The large band gap here can be related to the different atomic weights of Ge and C atom as the Ge atom is much heavier than a C atom.
The transverse acoustic modes (in- and out-of-plane) are not degenerate due to the structural anisotropy.
Additionally, while the LA and TA modes have a linear dispersion close to the $\Gamma$-point, the (ZA) phonons exhibit a quadratic dispersion around $\Gamma$-point.
The parabolic dispersion of ZA modes is a characteristic feature of layered materials. All properties of
the phonon band structure of a flat GeC monolayer are very well in agreement with the literature \cite{Guo_2019}.

In the presence of planar buckling, $\Delta \neq 0$, negative phonon frequencies are seen, especially for $\Delta \geq 0.3$~$\angstrom$ indicating dynamically unstable structures.
At $\Delta = 0.3$~$\angstrom$, only the ZA mode becomes negative in the vicinity of the $\Gamma$-point and
the negativity is further increased until all three acoustic modes become negative at the $\Gamma$-point, when $\Delta = 0.6$~$\angstrom$.
In addition, the phonon band gap is slightly increased in the presence of planar buckling affecting the thermal properties of the system.
\begin{figure}[htb]
	\centering
	\includegraphics[width=0.45\textwidth]{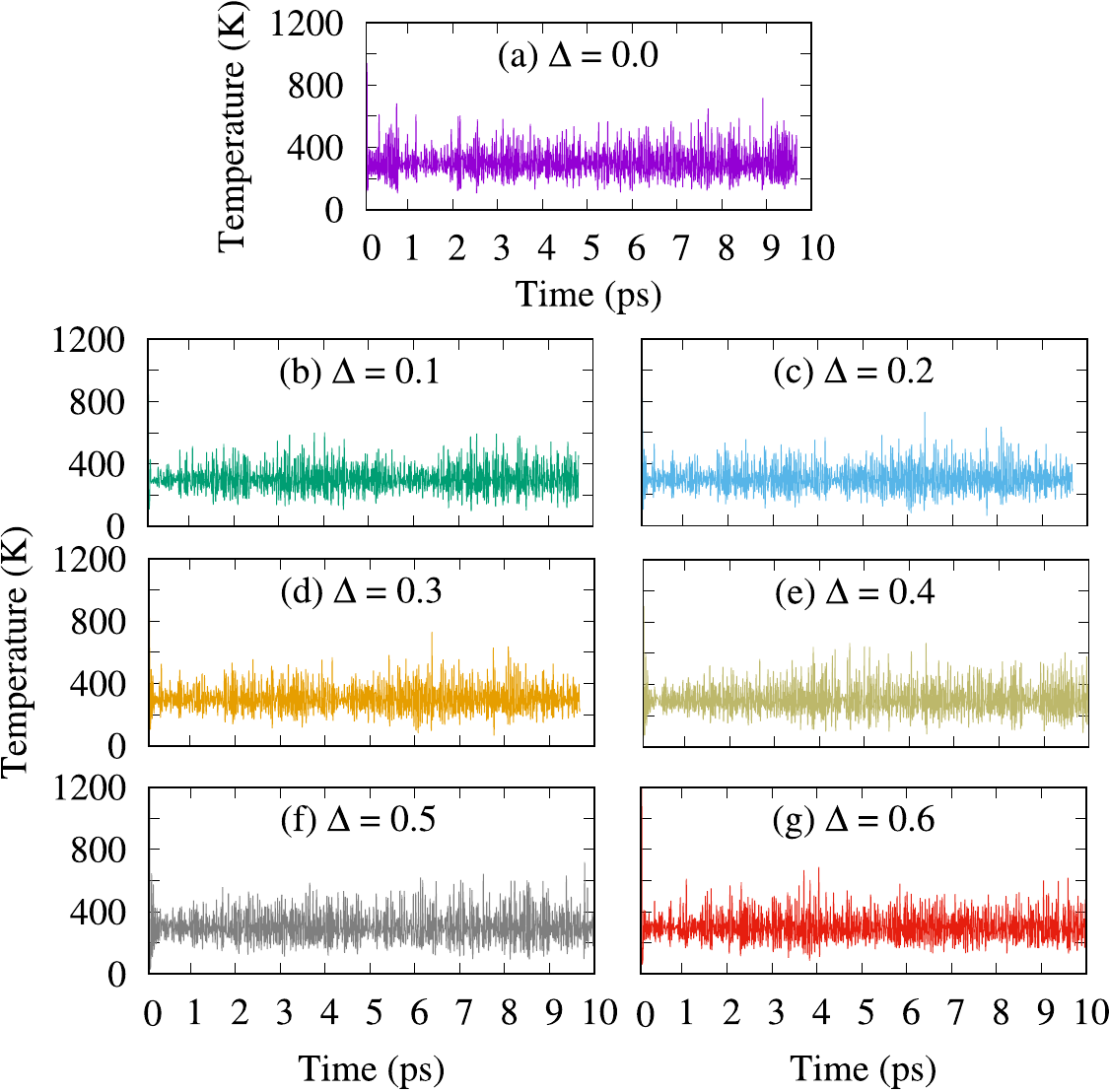}
	\caption{Temperature versus the AIMD simulation time steps at 300 K for optimized GeC monolayers with planar buckling, $\Delta = 0.0$ (a), $0.1$ (b), $0.2$ (c), $0.3$ (d), $0.4$ (e), $0.5$ (g), and $0.6$~$\angstrom$ (h).}
	\label{fig06}
\end{figure}
The thermal stability of a flat and buckled GeC is checked for approximately $10$ ps with a time step of $1.0$~fs as is presented in \fig{fig06}. The temperature curve of the pure and the buckled GeC monolayers neither displays large fluctuations in the temperature nor serious structure disruptions or bond breaking at $300$~K. This indicates that the pure and the buckled GeC monolayers are thermodynamically stable structures.

\begin{figure}[htb]
	\centering
	\includegraphics[width=0.45\textwidth]{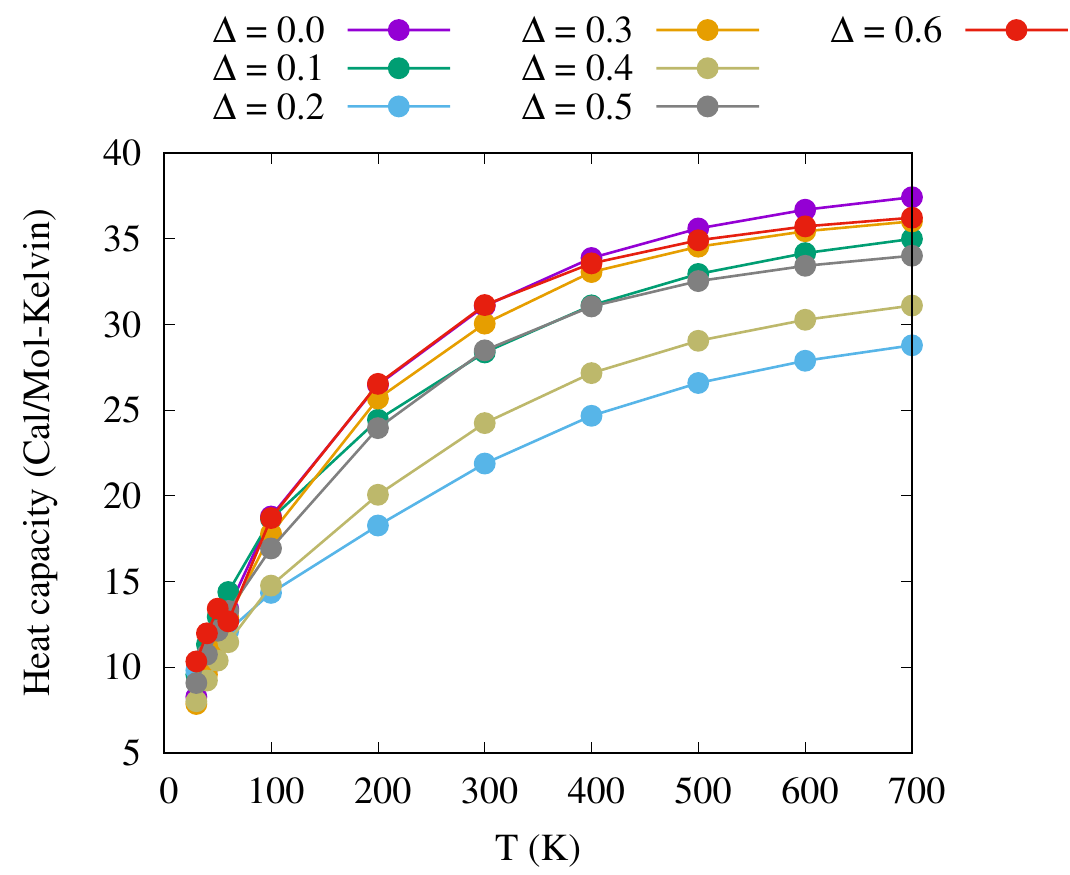}
	\caption{Heat capacity versus temperature for optimized GeC monolayers with planar buckling, $\Delta = 0.0$ (purple), $0.1$ (green), $0.2$ (blue), $0.3$ (orange), $0.4$ (olive), $0.5$ (gray), and $0.6$~$\angstrom$ (red).}
	\label{fig07}
\end{figure}

In \fig{fig07}, the heat capacity of a flat or buckled GeC monolayers is plotted.
The heat capacity is the ratio of the heat absorbed by a material to the temperature change. The heat capacity indicates not only the thermal energy stored within a body but also how quickly the body cools or heats.
As expected, the heat capacity of flat and buckled GeC monolayers increases with temperature, but the heat capacity is decreased with increasing planar buckling and the rate of the reduction is linear up to $\Delta = 0.2$~$\angstrom$. This is caused by the lifting of the crossings of ZO mode with TA and LA modes in the case of
$\Delta = 0.1$ and $0.2$~$\angstrom$, which has a significant effect on phonon scattering processes.
In the case of higher value of planar buckling, $\Delta > 0.3$~$\angstrom$, there is no linear relation between the heat capacity and the buckling $\Delta$. We can not rely on the results for the heat capacity as the negativity in the phonon modes is high in this range of planar buckling.

\begin{figure*}[htb]
	\centering
	\includegraphics[width=0.9\textwidth]{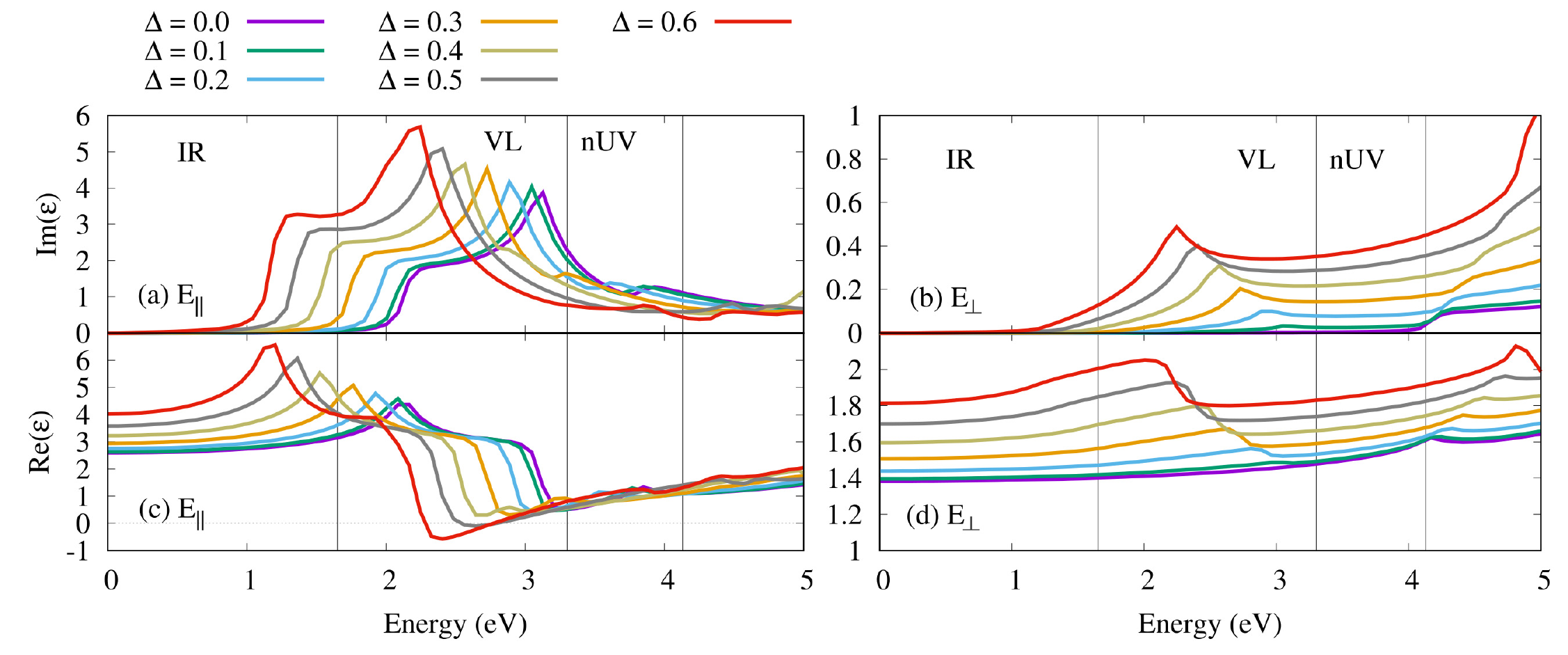}
	\caption{Imaginary, Im($\varepsilon$), (a,b) and real, Re($\varepsilon$), (c,d) parts of the dielectric function for the a GeC monolayer with buckling parameters, $\Delta = 0.0$ (purple), $0.1$ (green), $0.2$ (light blue), $0.3$ (orange), $0.4$ (light brown), $0.5$ (gray), and $0.6$~$\angstrom$ (red) in the case of E$_{\parallel}$ (left panel), and E$_{\perp}$ (right panel). The vertical black and red lines indicate different regions of the electromagnetic spectrum.}
	\label{fig08}
\end{figure*}

\subsection{Optical properties}
The optical properties in the first-principles calculations are defined by
the dielectric function. The complex dielectric function reflects the linear response of
the semiconductor materials to an electromagnetic radiation.
The random phase approximation (RPA) is utilized to calculate the complex dielectric function
with a very dense mesh grid, $100 \times 100 \times 1$, in the Brillouin zone to obtain accurate results \cite{ABDULLAH2022106835}. In addition, we calculate the refractive index, the electron energy loss spectra (EELS), and the optical conductivity.

We first consider the imaginary and the real parts of the dielectric function shown in \fig{fig08} for the parallel, E$_{\parallel}$, (a) and perpendicular, E$_{\perp}$, (b) polarization of the incoming electric field.
We focus on different regions of the electromagnetic spectrum that are highlighted via the vertical lines including the infrared, IR, ($0\text{-}1.65$~eV), the visible regime, VL, ($1.65\text{-}3.3$~eV), and the near ultraviolet, nUV, ($3.3\text{-}4.13$~eV).
For crystalline materials, the Im($\varepsilon$) is an important parameter that is associate with photon absorption. An anisotropy of this optical parameter can clearly be seen when comparing the results for the two polarization directions for both flat and planar buckled GeC monolayers.

In the absence of planar buckling, the main intensity peak in Im($\varepsilon$) (purple) for a flat GeC monolayer appears at $3.12$~eV in the case of $E_{\parallel}$. The peak starts at $2.25$~eV reflecting a transition over the optical band gap.
The main intensity peak is attributed to the electronic transition from C-p$_z$ in the valence band
to Ge-p$_z$ in the conduction band (see \fig{fig03}(a)). The main intensity peak is found in the VL region. Additionally, a very small peak (not significant) in Im($\varepsilon$) for a flat GeC is found at $4.25$~eV in the selected range of the electromagnetic spectrum for $E_{\perp}$.
In the presence of planar buckling, the main intensity peak in E$_{\parallel}$ is red shifted to a lower energy with higher intensity but it still remains in the VL region. This peak corresponds to the same transition from C-p$_z$ in the valence band to Ge-p$_z$ in the conduction band. Furthermore, a new peak is seen in the IR region at $1.28$~eV for the higher value of planer buckling, $\Delta > 0.4$~$\angstrom$.
The intensity of the new peak is increased with $\Delta$, and the new peak is red shifted with increasing $\Delta$. The new peak can be attributed to the electronic transition from C-p$_{x,y}$ in the valence band to Ge-s or Ge-p$_z$ in the conduction band (see \fig{fig03}(b)). This is caused by
the planar buckling effect redistributing the the orbital hybridization and the contribution of the density of states of the s- and p-orbitals.
We notice that the same applies for the $E_{\perp}$, but the new peak here appears in the VL region.
The zero value of Im($\varepsilon$) in the IR region for both directions indicates the semiconductor behavior of both the flat and the planar buckled GeC.

The real part of the dielectric function is an important physical parameter that can determine
the static dielectric constant (Re($\varepsilon^{(0)}$)) and the refractive index of the material. The real part of the dielectric function, Re($\varepsilon$), is presented in \fig{fig08}(c) and (d) for E$_{\parallel}$, and E$_{\perp}$, respectively. The static dielectric constant is defined as the real part of the dielectric function at zero value of the incident photon energy or frequency.
It can be seen that the static dielectric constant, Re($\varepsilon^{(0)}$), of a flat GeC is $2.95$ for E$_{\parallel}$ and $1.38$ for E$_{\perp}$ for zero frequency. The static dielectric constants for both polarization directions of the electric field agree with previous DFT calculations \cite{BEHZAD2019102306}.
The value of the static dielectric constant confirms that the flat GeC monolayer is a low dielectric material ($k<3.9$) and has semiconductor properties. By increasing the planar buckling, $\Delta$, the static dielectric function is increased to $4.02$ for E$_{\parallel}$ and $1.81$ for E$_{\perp}$ at $\Delta = 0.6$~$\angstrom$, respectively. The finite values of Re($\varepsilon^{(0)}$) show the semiconducting character of the monolayers.
This demonstrates that the planar buckling enhances the GeC monolayer from low to high dielectric material especially for E$_{\parallel}$.
Compared to a flat GeC, with the main peak of the real part at $2.12$ eV, it is red shifted for the buckled layers from the VL to the IR region for E$_{\parallel}$.

Another remarkable point for the real part of the dielectric function is the plasmon energy or frequency.
Plasmons can be determined by the real component of the dielectric function and are collective oscillations of the valence or the conduction electrons in a material. This is typically described by a material's complex dielectric function, in which the real component describes the electromagnetic wave transmission through the medium and the imaginary component describes single particle excitations or the damping of the collective mode, due to interband transitions \cite{raether2006excitation, egerton2011introduction}.
There are two elements indicating the presence of plasmon oscillations.
The real part of dielectric function changes sign or approaches a low value at the plasmon energy
and there is a sharp drop in the Im($\varepsilon$) curve to be seen as the Re$(\varepsilon)$ approaches zero \cite{Guan2015, doi:10.7566/JPSJ.91.074707}.
Therefore, the flat GeC monolayer shows indications for a plasmon mode at $3.2\text{-}3.5$~eV for E$_{\parallel}$ as there is a sharp drop in Im($\varepsilon$) and the Re($\varepsilon$) approaches zero.
The plasmon frequency for the buckled GeC monolayer is red shifted to the VL region at $\Delta = 0.5$ and $0.6$~$\angstrom$ for E$_{\parallel}$ as the  the real part of dielectric function drops and a minimum value in Im($\varepsilon$) is found. Additionally, no clear signs of plasmon oscillations is found for planar GeC for the E$_{\perp}$ polarization.

\begin{figure}[htb]
	\centering
	\includegraphics[width=0.45\textwidth]{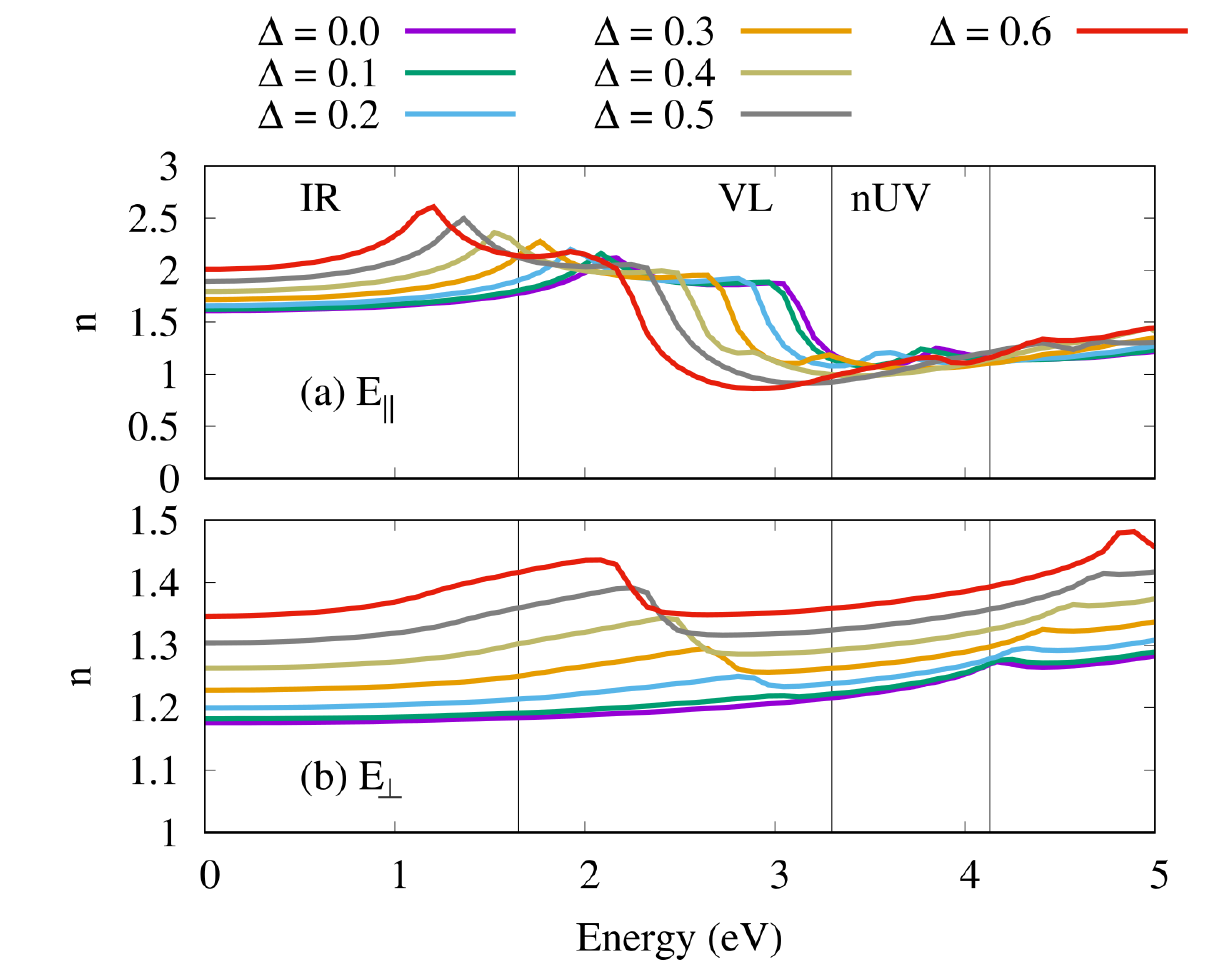}
	\caption{Refractive index, n($\varepsilon$), for the GeC monolayer with buckling parameter, $\Delta = 0.0$ (purple), $0.1$ (green), $0.2$ (light blue), $0.3$ (orange), $0.4$ (light brown), $0.5$ (gray), and $0.6$~$\angstrom$ (red) in the case of E$_{\parallel}$ (a), and E$_{\perp}$ (b). The vertical black and red lines indicate different regions of the electromagnetic spectrum.}
	\label{fig09}
\end{figure}

Another optical characteristic is the refractive index, $n(\varepsilon)$. In general, the behavior of $n(\varepsilon)$ follows qualitatively the characteristics of the real part of the dielectric function as presented in \fig{fig09} for both polarization directions of the incoming electric field. Large $n(\varepsilon)$ values display strong interactions between the incident photons and the valence electrons, which cause a reduction in the photon speed during the transmission.
The refractive index of a flat GeC monolayer at zero energy or frequency limit, $n(0)$, is $1.61$ and $1.17$, for E$_{\parallel}$ (a), and E$_{\perp}$ (b), respectively.
In the presence of planar buckling, $n(0)$ is increased to $2.0$ and $1.34$ at $\Delta = 0.6$~$\angstrom$ for E$_{\parallel}$ (a), and E$_{\perp}$ (b), respectively.

According to the refractive index curves in \fig{fig09}(a), the refractive index has its lowest value at $3.53$ eV (nUV region), which indicates weak interactions between the incident photons and the valence electrons. In this case, most of the light entering the flat GeC at this energy range passes through the monolayer, indicating a transparent material.
The minimum value of the refractive index is red shifted to the VL region, when the planar buckling is increased to $\Delta = 0.6$~$\angstrom$. It means that the transparency occurs at lower incident photon energy, when the planar buckling is considered. This is due to the weakness of the $\sigma\text{-}\sigma$ bond in the buckled GeC monolayer leading to more transparent material at lower photon energy. The maximum value of $n(\varepsilon)$ is seen in the IR region for E$_{\parallel}$, and in the VL region for the E$_{\perp}$. This demonstrates that a strong interaction between the incident photons and the valence electrons occurs in the IR and the VL regions for E$_{\parallel}$ and E$_{\perp}$, respectively.

\begin{figure}[htb]
	\centering
	\includegraphics[width=0.45\textwidth]{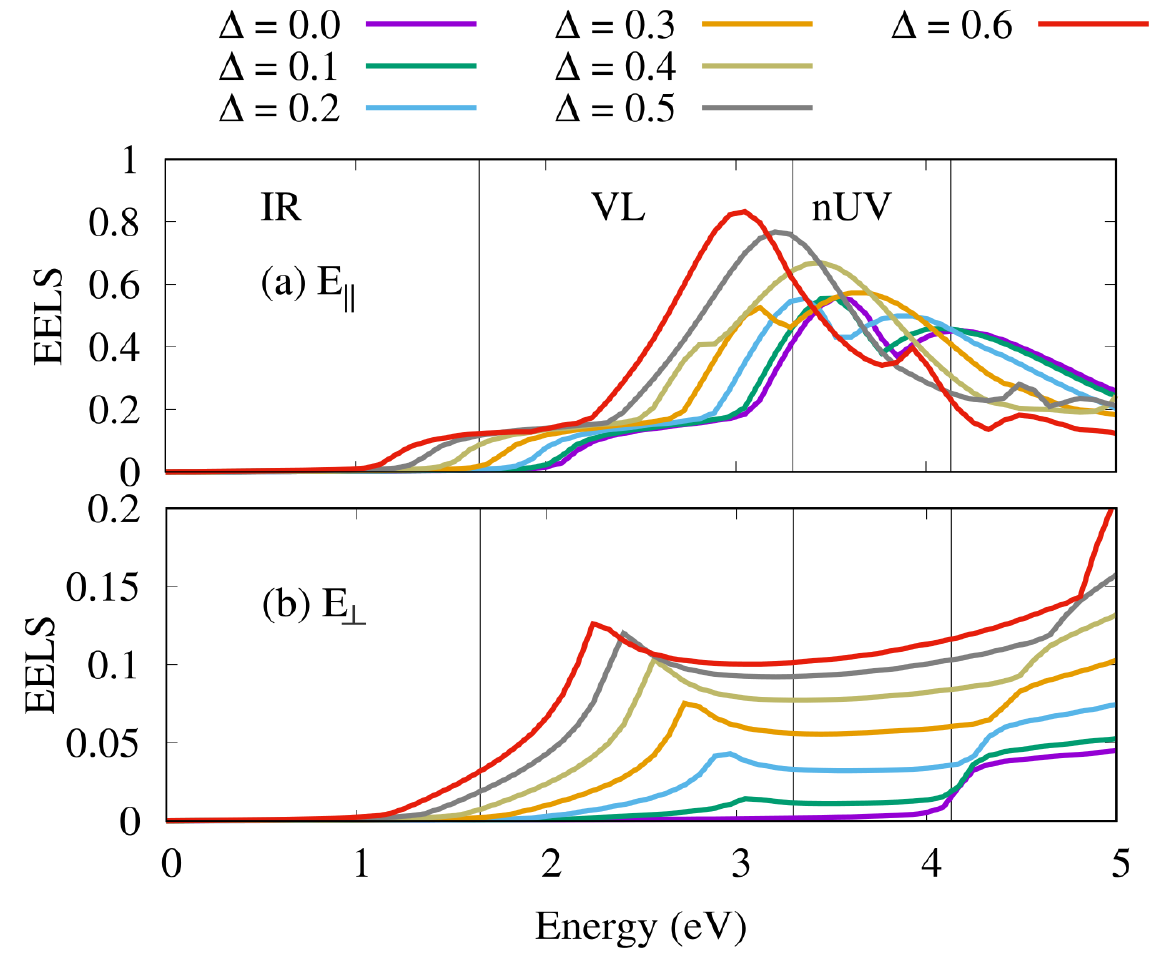}
	\caption{Electron energy loss spectra (EELS) for the GeC monolayer with buckling parameter, $\Delta = 0.0$ (purple), $0.1$ (green), $0.2$ (light blue), $0.3$ (orange), $0.4$ (light brown), $0.5$ (gray), and $0.6$~$\angstrom$ (red) in the case of E$_{\parallel}$ (a), and E$_{\perp}$ (b). The vertical black and red lines indicate different regions of the electromagnetic spectrum.}
	\label{fig10}
\end{figure}

Next, we discuss the electron energy loss spectra (EELS) presented in \fig{fig10}.
The EELS introduces the loss of energy between the incident photons and GeC monolayer interaction.
Again, the loss behavior for the two polarization directions is different. Each peak in the
loss curve demonstrates a significant energy loss for the incident photons.
In general, the reason for the EELS may be plasmonic oscillations or absorption or refraction of
light in matter.

\begin{figure}[htb]
	\centering
	\includegraphics[width=0.45\textwidth]{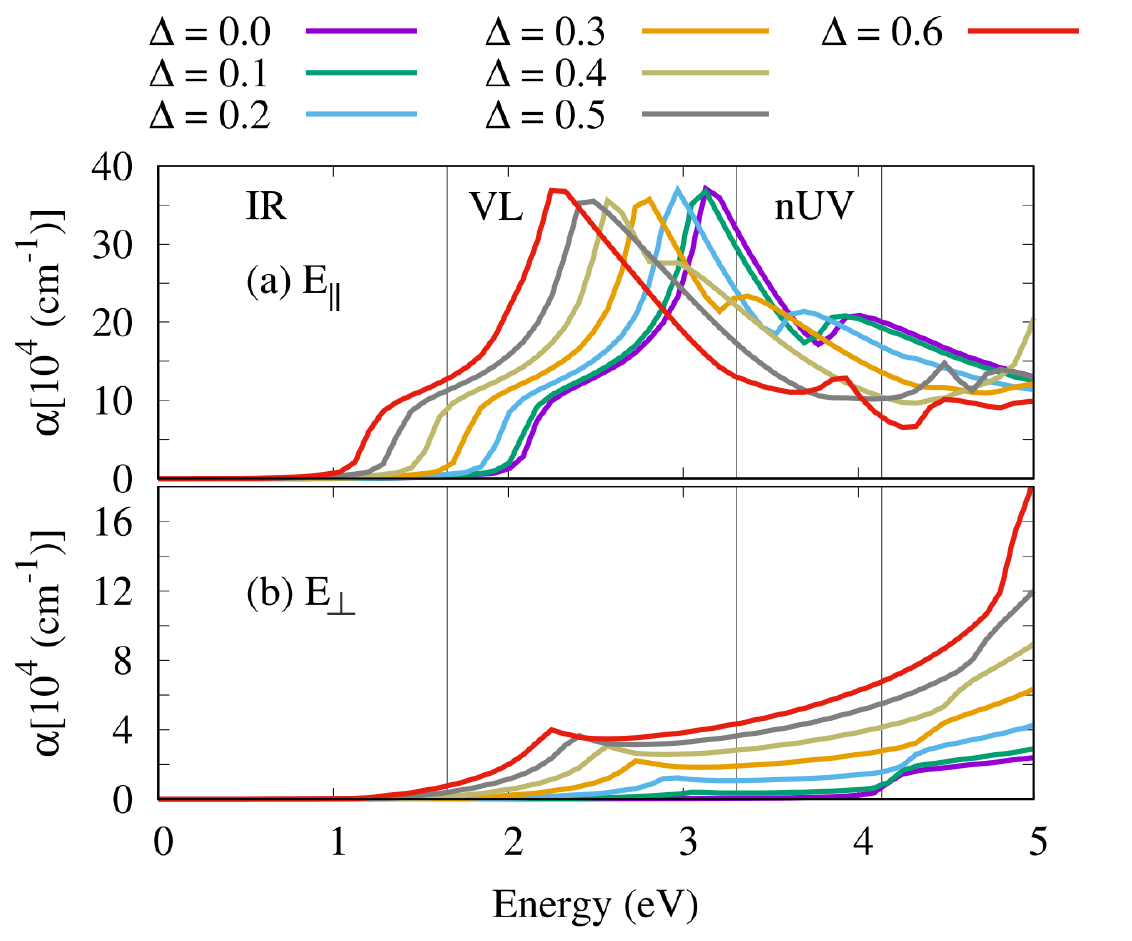}
	\caption{Absorption spectra ($\alpha$) for the GeC monolayer with buckling parameter, $\Delta = 0.0$ (purple), $0.1$ (green), $0.2$ (light blue), $0.3$ (orange), $0.4$ (light brown), $0.5$ (gray), and $0.6$~$\angstrom$ (red) in the case of E$_{\parallel}$ (a), and E$_{\perp}$ (b). The vertical black and red lines indicate different regions of the electromagnetic spectrum.}
	\label{fig11}
\end{figure}

The first significant peak of EELS is seen at $3.52$~eV (nUV region) for a flat GeC in the E$_{\parallel}$ direction. This peak is attributed to the plasmon oscillations shown in \fig{fig08}(a) and (c), whereas the effects of the refraction and the absorption do not play any role at this photon energy as is demonstrated in \fig{fig09}(a) and  \fig{fig11}(a), respectively.

Considering the buckled GeC, the maximum peak is red shifted with a higher energy loss at higher values of the planar buckling for both polarization directions. In the E$_{\parallel}$ direction, the maximum peak is found at $3.04$~eV in which the imaginary part of dielectric function has maximum value and the real part of dielectric function is zero. This loss peak is again due to damped plasmon oscillations.

The absorption coefficient ($\alpha$) for flat and buckled GeC monolayers for both the E$_{\parallel}$ (a), and the E$_{\perp}$ (b) is shown in \fig{fig11}. The zero value of $\alpha$ indicates that the GeC monolayer is not a good conductor at low photon energy for both polarization directions.
As the energy of the incident light increases, the absorption percentage increases in the VL region and the most intense peak is red shifted with increasing buckling, but remains in the VL region.
The absorption characteristics in the photon energy of $1.0\text{-}2.0$ eV in the E$_{\parallel}$ polarization direction can well be explained by the optical transition from the maximum of the valence band to the minimum of the conduction band in the band gap region.

\begin{figure}[htb]
	\centering
	\includegraphics[width=0.45\textwidth]{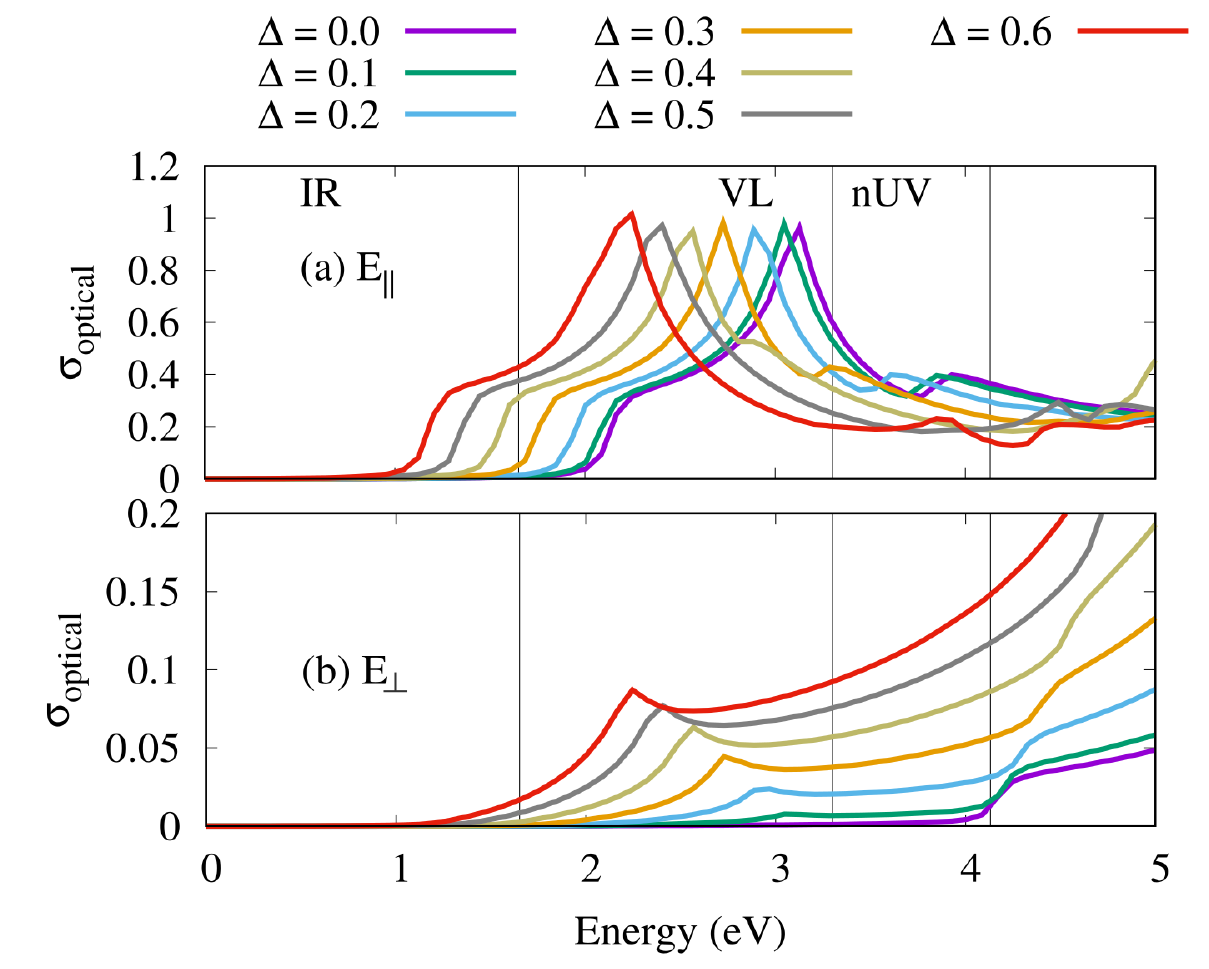}
	\caption{Optical conductivity ($\sigma_{\rm optical}$) for the GeC monolayer with buckling parameter, $\Delta = 0.0$ (purple), $0.1$ (green), $0.2$ (light blue), $0.3$ (orange), $0.4$ (light brown), $0.5$ (gray), and $0.6$~$\angstrom$ (red) in the case of E$_{\parallel}$ (a), and E$_{\perp}$ (b). The vertical black and red lines indicate different regions of the electromagnetic spectrum.}
	\label{fig12}
\end{figure}

Finally, the optical conductivity (real part), $\sigma_{\rm optical}$, for flat and buckled GeC monolayers is displayed in \fig{fig12}. The optical conductivity follows the same qualitative property of the absorption spectra. Wherever there is a peak in the absorption spectra at a specific photon energy,
a peak is found in the optical conductivity for almost the same photon energy.
This confirms that the optical conductivity is generated by the absorption of photons by the electrons in a specific energy level transferring the electrons to higher energy levels. This process can include both intraband and interband transitions.

\section{Conclusions}\label{conclusion}
In summary, we have used density functional theory to study the properties of a Germagraphene, GeC, monolayer by considering the buckling effects. The GGA-PBE functionals with full potential
augmented plane waves has been used in the calculations.
The findings for the electronic properties indicate that planar buckling results in a
tunable band gap, and the energy band gap decrease by increasing
planar buckling. This is due to redistribution of orbital hybridization and the contribution ratio of
the $s$- and the $p$-orbitals.
The band gap change can tune the thermal properties and improve the optical behavior in the visible light, the VL, range. The results for the optical properties demonstrate that all features of the optical spectra get
red-shifted to lower energy with increasing planar buckling.
The band gap reduction significantly enhances the absorption spectra in the VL region.
Consequently, the optical conductivity of buckled GeC is increased. Additionally, the heat capacity of flat and buckled GeC monolayers rises with temperature, but it falls with increasing planar buckling.

\section{Acknowledgment}
This work was financially supported by the University of Sulaimani and
the Research center of Komar University of Science and Technology.
The computations were performed on resources provided by the Division of Computational
Nanoscience at the University of Sulaimani.



\end{document}